\documentclass[final,5p,times,twocolumn,authoryear]{elsarticle}

\usepackage{url, xcolor, comment}
\usepackage{hyperref} 
\hypersetup{
        breaklinks=true,
	colorlinks=true,         
	linkcolor=Blue,          
	citecolor=Blue,
	urlcolor=Blue            
 }

\usepackage{amssymb}
\usepackage{lipsum}

\journal{}

\makeatletter
\def\ps@pprintTitle{%
 \let\@oddhead\@empty
 \let\@evenhead\@empty
 \let\@oddfoot\@empty
 \let\@evenfoot\@empty
}
\makeatother

\begin{document}

\begin{frontmatter}

\title{The Spectre of Underdetermination in Modern Cosmology}

\author[first]{Pedro G. Ferreira}
\author[first,second]{William J. Wolf}
\author[second]{James Read}
\affiliation[first]{Astrophysics, University of Oxford, DWB, Keble Road, Oxford OX1 3RH, United Kingdom}
\affiliation[second]{Faculty of Philosophy, University of Oxford}
\date{\today}

\begin{abstract}

\noindent The scientific status of physical cosmology has been the subject of philosophical debate ever since detailed mathematical models of the Universe emerged from Einstein's general theory of relativity. 
Such debates have revolved around whether and to what extent cosmology meets established demarcation criteria for a discipline to be scientific, as well as determining how to best characterize cosmology as a science, given the unique challenges and limitations faced by a discipline which aims to study the origin, composition, and fate of the Universe itself. The present article revisits, in light of the dramatic progress in cosmology in recent decades, an earlier debate held in the 1950s between Herman Bondi and Gerald Whitrow regarding the scientific status of cosmology. We analyse cosmology's transition from an emerging science to a cornerstone of modern physics, highlighting its empirical successes in establishing the $\Lambda$-Cold Dark Matter ($\Lambda$CDM) model and in its delivery of various successful novel predictions. 
Despite this remarkable scientific success and progress, we argue that modern cosmology faces a further profound challenge: the permanent underdetermination of the microphysical nature of its exotic energy components: inflation, dark matter, and dark energy. Drawing historical parallels with the role of spectroscopy in revealing the microphysical nature of atomic physics, we argue that the epistemic barriers obstructing us from ascertaining the microphysical nature of these exotic energy components are significant, in turn casting doubt upon whether cosmology can ever transcend these particular epistemic challenges. We conclude by reflecting on the prospects for future breakthroughs and/or non-empirical arguments which could decide this issue conclusively.

\end{abstract}

\end{frontmatter}




\section{The Bondi--Whitrow debate}

In 1953, an article penned by Herman Bondi and Gerald Whithrow appeared in \emph{The British Journal for the Philosophy of Science} with the provocative title ``Is Physical Cosmology a Science?'' \citep{Whitrow1953-WHIIPC}. The article was catalysed by Whithrow's earlier review of Bondi's monograph, {\it Cosmology}. There, \cite{bondi1960cosmology} declared that ``The aim of this book is to present cosmology as a branch of physics". He then set out to describe the mathematical framework for cosmology developed over the previous half century, the (then) current observations supporting that framework, and the diversity of models of the Universe which could be accommodated within it. 

Bondi himself was a fervent advocate of the Steady State Universe. This model posited that the Universe had always existed, was expanding at a constant rate, and possessed a constant energy density that was regenerated by an elusive source that permeated all of space. This was in contrast to a competitor model which assumed that the Universe had been expanding and dynamically evolving in time---but from some initial primordial beginning or singularity, perjoratively dubbed ``the Big Bang" by the supporters of the Steady State Universe.

Whithrow took it upon himself to scrutinize Bondi's case that cosmology was a science. He formulated the challenge as ``Is Physical Cosmology a Science?''---and he made the case that the answer was ``No''. 
Underlying Whithrow's case was the claim that ``opinion and personal taste''---which play a crucial role in philosophy, arts and humanities---should be absent entirely from the practice of science. While they might play a role ``at the periphery of knowledge'', they should be absent entirely from the main developments. And yet, he claimed, they seemed to be an integral part of the practice of cosmology at the time. For example, different definitions of the ``cosmological principle'' were adopted by supporters of different theories based on aesthetic grounds.\footnote{In its most basic form, the cosmological principle can be taken to mean that the Universe and its properties are the same everywhere. It is in the meaning of ``everywhere'' that the different camps disagree. For the Steady State apologists, it means all of space and time while for the proponents of the Big Bang, it is restricted to space alone.} Furthermore, if one stepped back, one might argue that such a situation hadn't changed in cosmology over the past few millennia. While the details of the cosmological models had changed since ancient Mesopotamia and the recent centuries had adopted mathematics, the same uncertainty prevailed. At the time of writing, it was, indeed, a matter of opinion whether one believed in the Steady State or the Big Bang cosmology.

Bondi responded that physical cosmology was a science because it asked and answered questions using the tools of science. As he stated, ``the most characteristic feature of any science is that it confines its attention to the establishment of connections between existing results of experiment and observation, and to the forecasting of new ones. The chief claim that can be made for any scientific theory is that it fits the facts, and forecasts (or has forecast correctly) the results of new 
experiments.'' Bondi argued that the absence of personal opinion or ``unanimity'' only occurred when the science was already in an advanced state. He invoked other examples (such as ``virus research'') where there was no unanimity but for which, nevertheless, there was no doubt they were scientific endeavors. As he put it, ``What is essential is only that there should be unanimity about the means of deciding what is correct and what is incorrect, that the yardstick of experiment should be universally accepted'' \citep[p.~278]{Whitrow1953-WHIIPC}.\footnote{Withrow exhibits an idealised conception of science as value-free. Bondi's response anticipates the literature on the interaction between the value-ladenness of science and its objectivity: ``But objective scientific results do not, or so the argument goes, depend on researchers’ personal preferences or experiences---they are the result of a process where individual biases are gradually filtered out and replaced by agreed upon evidence.'' \citep{sep-scientific-objectivity}.}

Bondi's views were sensible and rooted in scientific practice. Philosophers will recognize that he was operating within a Popperian philosophy of science, adopting his famous demarcation criteria of falsifiability:  
``statements or systems of statements, in order to be ranked as scientific, must be capable of conflicting with possible, or conceivable observations'' \citep[p.~39]{Popper1962-POPCAR-5}.\footnote{Interestingly, while Bondi was thinking about science in a very Popperian way, Popper's own views on cosmology seemed to be closer to those of Whitrow \citep{Kragh2021}.} At the time of the Bondi--Whitrow article, it was clear that cosmology was at least partially delivering on its promise as a scientific endeavor understood in this way, as Hubble's empirical measurements of the recession of galaxies had delivered a fatal blow to Einstein's static model of the Universe and set cosmologists onto the path of expanding models like the Steady State or the Big Bang.

However, at that time there had been only incremental and inconclusive progress in measuring the expansion of the Universe. While the uncertainties in the measurement of the Hubble constant had decreased substantially, the actual values nevertheless depended greatly on the methods being used, leading to wildly inconsistent results. Furthermore, there were very few attempts at measuring other properties of the Universe which might help to shed light on, and ajudicate between, the different rival models. The situation ten years after the Bondi--Whithrow debate is described succinctly by Jim Peebles. As he put it, in the early 1960s, cosmology was ``a limited subject---a subject, as it used to be advertised, with two or three numbers''; moreover, he added, ``science with two or three numbers always seemed to me to be pretty dismal'' \citep{peebles1988interview}. Clearly, the mere existence of some potentially falsifiable claims does not by itself guarantee that a science will be something more than ``dismal'' or that it will ever progress towards the kind of unanimity that Whitrow demanded.\footnote{But even Whitrow's conviction that science's defining feature is a forged consensus amongst experts also leaves something to be desired as it does not seem to be able to account for the stream of successive revolutions and paradigm shifts that have consistently upset unanimity across so many unobjectionably ``scientific'' disciplines \citep{Kuhn1962-KUHTSO-3}.}

\section{Physical cosmology as a successful empirical science}

There are many illuminating ways in which one might characterise science. Recently, \cite{Peebles:2024jdt} penned an analysis that, in his view, captures the implicit operative philosophical intuitions held by the vast majority of working physicists: a ``physicists' philosophy of physics''. 
One can distill two core ideas from Peebles' starting assumptions: (1) an unabashed and unapologetic commitment to scientific realism, by which he means that physicists subscribe to the existence of a mind-independent reality and share a conviction that this mind-independent reality can be accurately captured by ever better approximations (i.e., our scientific theories), and (2) placing a premium value on the successful empirical confirmation of predictions as being the primary indicator of the approximate truth of our theories. 

One particular quality that has been celebrated consistently in the main accounts of scientific progress (despite their many differences), and that has an unmistakable bearing on both (1) and (2), is the idea that successful science should deliver (and confirm) \textit{novel} empirical predictions \citep{Popper1935-POPTLO-7, Kuhn1962-KUHTSO-3, Lakatos1978-LAKTMO}. While predictive novelty has traditionally been construed in a \textit{temporal} sense, meaning that the predicted and subsequently confirmed phenomena was not known beforehand, there are alternative conceptions such as \textit{heuristic}/\textit{use} novelty, which roughly means that a prediction can still count as novel even if the phenomena were known beforehand, as long as they were not used explicitly in the construction of the theory \citep{Zahar1973-ZAHWDE, Gardner1982-GARPNF, worrall1985scientific}.\footnote{I.e., the theory was not specifically designed to accommodate the phenomena. For example, the perihelion precession of Mercury was known before the discovery of general relativity, but Einstein's use of the theory to predict/explain this was still hailed as a crucial and novel epistemic achievement of the theory \citep{Zahar1973-ZAHWDE, Worrall2014-WORPAA-2}.} In any case, the exact construal of novelty will not matter much for what follows as the particular achievements highlighted can be understood to succeed on either (or both) of these primary accounts.  

Novel predictions are widely considered to be the gold standard of empirical success and to hold special epistemic value within science. Some reasons for this include the following. It is a contingent fact that we have successfully used scientific theories to discover and/or understand ``novel'' facts about the world. This strongly suggests that such theories are latching onto an approximately true description of the mind-independent reality that is at the heart of the physicists' philosophy of science, otherwise such success would be unexplained and seemingly miraculous \citep{McMullin1984-MCMACF}. Additionally, it is nearly always possible to accommodate empirical data with \emph{ad hoc} hypotheses and tinkering, which arguably only reflect theorists' ingenuity in crafting a theory to fit the data. Such accommodation does not, by itself, provide a strong evidential basis for the theory (besides meeting the minimum threshold of empirical adequacy). Novel predictions, on the other hand, provide independent evidence for the rationally warranted acceptance or confirmation of the theory \citep{worrall1985scientific}. Furthermore, novel predictions play a crucial role in the creation of new evidence and uncovering new evidence--theory relations, provide assurances against over-fitting, and directly facilitate the further production of empirically successful scientific theories \citep{Douglas2013-DOUSOT}.

With this in mind, it will soon become apparent that cosmology has succeeded as an empirical science not only according to the standards of unanimity (Whitrow) and falsifiability (Bondi), but also in the sense that its empirical success has consistently proved to be fruitful and novel.

The status of physical cosmology changed dramatically from the 1960s onwards. Here we will focus on three main episodes:
    (i) the confirmation of the Big Bang model over the Steady State model,
    (ii) an intermediate phase with a proliferation of various Big Bang cosmological models, during which evidence for $\Lambda$CDM began to appear, and
    (iii) the confirmation of the $\Lambda$CDM model as the standard model of Big Bang cosmology.

While the primary focus of experimental cosmology up to the 1960s had been to determine accurate measurements of the expansion rate of the Universe---those ``two or three numbers'' to which Peebles referred---the wealth of new data at radio frequencies would greatly widen astronomers' view of the Universe. Notably, it would conclusively settle the debate between the rival Steady State and Big Bang theories.

There were crucial theoretical insights that preceded these developments, 
as theorists had worked out several consequences or predictions that would follow naturally from a universe described by a Big Bang model. These include (see \cite{ferreira2014perfect, Peebles:2022bya} for the historical details): the expectation of larger numbers of bright radio sources in the early Universe due to the higher densities and more active star formation processes at these early times; the existence of relic radiation, left over from the hot initial state in the early Universe; the realization that the Bang Big nucleosythesis (BBN) of light elements implied a very large relic abundance of Helium relative to what would be expected if stellar processes were the primary Helium source (as in the Steady State model).

By the mid 1960s, a key set of observations played a decisive role in physicists' weighing of the Big Bang and Steady State models.  Measurements of the number density of radio sources and its redshift dependence were shown to be consistent with the Big Bang theory. The relic radiation was discovered in the form of the cosmic microwave background. This smoking gun of the hot Big Bang Universe was found at a temperature of $\sim3K$, in strikingly close agreement with theoretical predictions made in the prior decade. And finally, various spectroscopic observations of stars and galaxies confirmed that there was a large Helium abundance in the Universe, in agreement with earlier BBN calculations. While there were attempts to fix the Steady State model, all of these observations were grossly inconsistent with it. 
Thus, the Steady State model was jettisoned rapidly from the canon of viable cosmological models by the majority of practicing astronomers and physicists.

The adoption of the Big Bang model precipitated rapid progress in cosmology\footnote{The empirical verification of the cosmological principle is still a topic of research today. Notable attempts at devising methods for doing so have been presented, or discussed in \citet{Clarkson:2007pz,Maartens:2011yx,February:2012fp}.}. From the late 1960s onwards, slowly but surely a fully fleshed-out mathematical theory of the Universe was developed that could be used to accurately describe not only its expansion, but also the inhomogeneous large-scale structure of spacetime mapped out by galaxies and clusters. P.J. Peebles, M. Rees, R. Sachs, J. Silk, Y. Zel'dovich, and collaborators \citep{SachsWolfe1967,1968ApJ...151..459S,1968Natur.217..511R,1970ApJ...162..815P,1978SvA....22..523D,1981ApJ...243...14W,1982ApJ...263L...1P} built up a mathematical framework based on Einstein's theory of general relativity, statistical mechanics, and radiative transport which allowed them to make statistical predictions of how gravitational collapse would evolve on scales where linear perturbation theory was valid.  By the 1980s, the numerical techniques were such that M. Davis, G. Efstathiou, C. Frenk, and S. White \citep{1985ApJ...292..371D} were able to simulate model universes, now including the non-linear regime of gravitational collapse, and predict how galaxies would assemble, evolve and coalesce to form what is now called the cosmic web. In parallel, J. Bond and G. Efstathiou \citep{1984ApJ...285L..45B,1987MNRAS.226..655B}
produced the first modern predictions for how the inhomogeneities would affect the cosmic microwave background. On the observational front, the first large scale surveys of galaxies were assembled \citep{1977ApJ...217..385G,1982ApJ...253..423D,1988IAUS..130..151M,1990ApJ...361...49S}, as were the first attempts at measuring anisotropies in the cosmic microwave radiation \citep{2009fbb..book.....P}. 

Yet, cosmologists were still far from attaining unanimity. \citet[Ch.~8]{Peebles:2022bya} describes the state of the field as late as the early-to-mid 1990s as ``cautious'', with physicists seriously considering a variety of Big Bang models, including Einstein-De Sitter models, mixed cold-hot dark matter models, and others, as well as the $\Lambda$-Cold Dark Matter ($\Lambda$CDM) model. However, by this point, there was empirical evidence that was pointing towards $\Lambda$CDM, whose ingredients we'll now briefly recall. 

The first ingredient of the $\Lambda$CDM model is known as \textit{inflation}. The Universe presumably must have undergone \textit{some} process at early times that would lead to development of perturbations/inhomogeneities that would seed structure in the later Universe; the usual hot Big Bang expansion did not provide any explanation for this and so any perturbative features had to be encoded in the initial state in a seemingly \emph{ad hoc} manner. A solution that has great appeal in the cosmological community is that the Universe underwent a period of accelerated expansion, driven by a fundamental scalar field known as the \textit{inflaton}. This inflationary period would stretch quantum fluctuations from microscopic scales to macroscopic scales, leaving a set of classical, approximately scale-invariant perturbations, as well as producing a spatially flat universe \citep{Guth:1980zm, Starobinsky:1980te,Linde:1981mu,Mukhanov:1981xt,Albrecht:1982wi, Bardeen:1983qw, Hawking:1982cz}.\footnote{Neither the CMB anisotropies nor the geometry of the Universe had been measured accurately in the 1980s. There were some very loose constraints, but reliable detections/measurements of these quantities took over a decade to come to fruition. See \cite{WolfDuerr} for an analysis of novelty in the context of these inflationary predictions.}

The second feature is that galaxies require the existence of a large reservoir of clustering mass that can't be accounted for by the current census on baryons. This (cold) {\it Dark Matter} (CDM) does not interact with light and helps keep galaxies together, preventing them from being ripped apart by the motions of the stars in their gravitational field. Crucially, cosmological evidence 
for the existence of CDM became very compelling when experimental upper bounds on anisotropies in the CMB indicated that there was not enough baryonic matter to account for the observed structure in the later Universe \citep{Peebles:1982ff}.

The final feature is the accelerated expansion at late times which can be accurately explained by the presence of some form of energy which doesn't cluster or interact with light. The simplest proposal for this {\it Dark Energy} is the cosmological constant $\Lambda$, but the possibility that it might be some other form of energy with peculiar statistical and thermodynamic properties can certainly be considered. The case had been made tentatively for a cosmological constant due to the fact that a purely matter-dominated universe was too young to harbour  globular clusters \citep{Turner:1984nf}. \cite{Efstathiou:1990xe} were among the first to make strong empirical arguments that CDM alone could not account for the distribution of galaxies seen in the galaxy correlation function, but that introducing $\Lambda$ as the dominant energy component at late times could explain these more intricate details of structure formation. \cite{ostriker_steinhardt_nature} also used a variety of observational constraints to argue that there was substantial evidence for dark energy.

The $\Lambda$CDM model\footnote{Throughout we will use ``$\Lambda$CDM" to denote the cosmological model with dark energy which is not necessarily restricted to the cosmological constant, $\Lambda$.} thus came into focus and was pieced together as a plausible model of the Universe using a somewhat intricate cocktail of emerging empirical evidence and explanatory reasoning. However, a slew of novel predictions would soon be confirmed from the mid 1990s onwards. To begin, CMB experiments such as
 COBE \citep{Smoot:1998jt}, Boomerang \citep{Boomerang:2000efg}, WMAP \citep{Komatsu:2014ioa}, and Planck \citep{Planck:2018vyg} first detected and then mapped out in exquisite detail the anisotropies in CMB, confirming the inflationary predictions of approximately scale invariant fluctuations in a spatially flat universe. The detailed measurements of the CMB angular power spectrum across a wide range of angular scales by WMAP and Planck confirmed earlier predictions from \cite{Bond:1984fp} regarding the ratios of peaks in the CMB power spectrum in the presence of significant amounts of CDM. In parallel, measurements of the distances and redshifts of high redshift supernovae indicated that the Universe's rate of expansion is accelerating and thus is indeed dominated by a large dark energy component \citep{SupernovaCosmologyProject:1998vns,SupernovaSearchTeam:1998fmf}.

These and many subsequent results have cemented $\Lambda$CDM as the standard model of cosmology. The main ingredients---inflation, dark matter, and dark energy---are incorporated into the mathematical model at the phenomenological level as they represent additional sources of energy with particularly desirable properties which allow the mathematical model to work. Yet, they give us a perfectly complete and predictive model of the Universe which allows us to constrain cosmological parameters---such as the Hubble constant, the density of matter and radiation density, the amplitude and initial morphology of perturbations in the Universe and even the total mass of the neutrino species---to sub-percentage precision. 

Taking stock, it's fair to say that cosmology has lived up to Bondi's ambitions for the field which he promoted during its nascency, and forged the unanimity which Whitrow demanded of any successful empirical science. Even more than this, cosmology has succeeded by the highest epistemic standards of science, consistently delivering and confirming novel predictions which have enhanced our understanding of the Universe.

\section{The rise of particle cosmology}



While reflecting on the nature and structure of scientific theories, \citet[pp.\ 361--2]{Einstein1936} distinguished between two kinds of physics. One kind he referred to as ``phenomenological'' physics,  which ``makes as much use as possible of concepts which are close to experience but, for this reason, has to give up, to a large extent, unity in the foundations''. Accordingly, he noted that phenomenological theories like both thermodynamics and the physics of continuous media describing heat phenomena ``can only have an approximate significance''. The other kind of physics, which at various points Einstein labelled as ``constructive'',  ``mechanical'', or ``atomistic'', is exemplified by the kinetic theory of gases. This kind of physics is primarily focused on the detailed microphysical underpinnings of the phenomena. The kinetic theory of gases ``gave a mechanical interpretation of the thermodynamic ideas and laws and led to the discovery or the limit of applicability of the notions and laws of the classical theory of heat'' and ``far surpassed
phenomenological physics as regards the logical unity of its foundations, [and] produced, moreover, definite values or the true magnitudes or atoms and molecules''.\footnote{Roughly, the distinction drawn by \cite{Einstein1919} between ``principle theories'' and ``constructive theories'' tracks the distinction between phenomenological theories of physics and theories which seek to erect microphysical models. In this article, we've largely elided the principle/constructive terminology, because the distinction is by now well-explored, and has delicate connotations which we don't wish to detract from the main points we seek to make in the body of this article. For further discussion of the principle/constructive distinction, see e.g.\ \cite{Acuna2016-ACUMSA, VanCamp2011-VANPTC-2, Brown2005-RBRPRS-2}.}

It is, however, important to note that this ``lack of unity'' should not be taken to imply that there is 
anything necessarily deficient about phenomenological physics. Although it is of course sensible to prefer a microphysical theory (if it can be found) on the grounds that it provides a complete ontology and understanding of the phenomena at a more fundamental level, it is quite often necessary to deploy phenomenological physics in order to first grasp the phenomena and to avoid ``becoming hopelessly lost in this line of thought (atomistic)'' \cite[p.~360]{Einstein1936}. Indeed, it seems quite natural to think that optimal science displays a reflective equilibrium between both approaches. As new data is collected which can allow physicists to better grasp the phenomena and distill useful principles and phenomenological descriptions, they can also build microphysical models which can fit the constraints imposed by that data, and eventually move towards attaining a detailed understanding of the microphysical underpinnings of the phenomena.

One can clearly see this process play out in different episodes in the history of cosmology sketched above. 
For example, it was clear by the 1980s that there was cosmological evidence for \textit{some} kind of weakly interacting massive contribution to the energy budget, but cosmologists were not expecting these observations from detailed models of the dynamics of galaxies or matter perturbations using known microphysics. Thus, these observations act as constraints on future microphysical model building, but as we have seen, integrating this phenomenologically distilled concept into our cosmological modeling (i.e.,\ there exists some form of significant, but weakly interacting mass) has still proved enormously fruitful for cosmology even without an accepted microphysical theory of the phenomena.

While this was playing out, one can also see examples where a different role for cosmology was emerging; a cosmology that could be a laboratory for fundamental physics with the ability to probe the properties of particles and fields which were inaccessible by other means. The idea that the very early Universe could be seen as a laboratory for high energy physics had been  considered sporadically since the first relativistic cosmological models were proposed.\footnote{Lem\^aitre  coined the term ``primordial atom'' or ``primordial egg'' to describe the microscopic processes which could have occurred at the earliest time \citep{1931Natur.127..706L}
. Indeed, he went far enough to discuss the cosmological constant as the energy density of the vacuum \citep{1934PNAS...20...12L}. Zel'dovich revisited Lem\^aitre's idea of the vacuum energy and discussed in detail how to relate it to  quantum field theory \citep{Zeldovich:1968ehl}.} We have already seen the first successful step in this direction with the work (of Gamow and his collaborators) in developing BBN in the late 1940s, deriving (and in this case \textit{predicting}) the elemental composition of the Universe with a detailed microphysical model of the nuclear and thermal interactions between the known baryonic particles. Throughout the 1960s and 1970s there were forays into the interplay between particle physics and cosmology.  
But it was most notably in the early 1980s that the role of fundamental fields in the early Universe became an active field of research.\footnote{This is also apparent as the Particle Data Group's review of particle physics began including a dedicated section on the Big Bang model and cosmological constraints in the 1986 edition \citep{AguilarBenitez:1986sb}.} With the resounding success of field theory in establishing the standard model of particle physics, the idea that fundamental fields could not only play an important role in the history of the Universe, but also that cosmological measurements might be used to learn about fundamental fields, took hold. 

Two decisive developments entrenched this link between particle physics and cosmology. The first had to do with attempts to identify the fundamental nature of dark matter. The idea that massive neutrinos could constitute dark matter was rejected rapidly when it was shown that, in such a scenario, all structure would be erased on megaparsec scales and below, precluding the possibility of the formation of galaxies. With the rise of grand unified theories and, specifically, the role of supersymmetry, the possibility that a weakly interacting superpartner could constitute dark matter took hold, fueling a systematic programme for direct detection that has survived until today. Thus, these developments can be seen as an attempt to discover the microphysical model underlying the empirical regularities that suggested the existence of this type of matter.

Arguably even more significant was the aforementioned discovery that a phase of inflation-driven expansion in the early Universe could not only solve a number of fine-tuning problems in the cosmological model, but also lead to plausible initial conditions for large scale structure. While the idea of exploring the properties of scalar fields in an expanding universe wasn't particularly new,\footnote{A few notable examples are how a dynamical, cosmological, spacetime would affect quantum scalar fields (leading to particle production) \citep{Parker:1969au} and how these fields might affect the expansion rate of the Universe (in particular, at the initial singularity) \citep{Zeldovich:1971mw}.} with inflation these fundamental fields became a core ingredient to the Universe's dynamics.
Over the following decade a new industry developed: one of constructing models, making predictions for how they would affect large scale structure and the early Universe, and more generally, consolidating the idea that the Universe could be used as an immense laboratory for particle physics.\footnote{Textbooks delving into the minutiae of how to extract information about the early Universe were written, such as {\it The Early Universe} \citep{KolbTurner1994}. Moreover, memorable conferences and workshops were held: {\it The Very Early Universe} held in 1982 in Cambridge and {\it Inner Space and Outer Space} in 1984 in Fermilab are two examples.}

The discovery of anisotropies by the COBE satellite \citep{Smoot:1998jt}, and the detailed investigations of their properties by successor satellites like WMAP \citep{WMAP:2006bqn} and Planck \citep{Planck:2015fie}, proved to be a pivotal event for solidifying this link between fundamental physics and cosmology. Measurements of these anisotropies showed that the primordial power spectrum of density fluctuations, which would have been generated in the inflationary era, is \textit{almost}---but not exactly---scale invariant. The current constraint on the scalar spectral index is $n_S=0.9649\pm .0042$, where $n_s=1$ corresponds to scale invariance \citep{Planck:2018vyg}. Remarkably, this is what one would in general expect from an inflationary regime described by a fundamental scalar field evolving in a potential. If it had been exactly scale invariant, then there would have been a problem in explaining how inflation ended.\footnote{This is because $n_s$ is pushed slightly away from $1$ if the scalar field is actually evolving as opposed to remaining static.} Yet, this value of the scalar spectral index corresponds both to what we would expect if an inflationary phase did occur and what is needed for successful structure formation. One of the most remarkable aspects of this story is that it was almost entirely theoretical developments that took the lead; predicting and explaining how the large scale structure seen today emerged from microphysical processes: the quantum fluctuations of a fundamental scalar field.\footnote{See \cite{Smeenk2018-SMEIAT} for a historical overview.}

In light of this dizzying success, the hope was that, with further precise measurements, it would be possible to actually pin down the exact details of the microphysical models. In other words, the goal of this strand of research became to reproduce the successes of particle physics in the 1960s and 70s and to determine the fundamental action for inflaton, dark matter and dark energy. Cosmology would then enable us to find the complete theory of the Universe. As of now, these efforts persist with respect to all three main ingredients of $\Lambda$CDM.

Inflationary cosmology focuses on the possibility of mapping out, in detail, the inflationary potential which governs the dynamics of the inflaton field. The scalar spectral index is not the only constraint that sheds light on fundamental physics during the inflationary regime. Current CMB constraints place stringent upper bounds on the number of gravitational waves generated during the inflationary phase. In the same way that quantum ripples in the inflaton field could seed large scale structure, quantum ripples in spacetime itself could be stretched and amplified to large scales to form a bath of classical gravitational waves. A new generation of experiments is actively searching for the signatures of these primordial gravitational waves, so far with little success. Yet the upper bounds on primordial gravitational waves are already allowing us to eliminate some of the simplest models of inflation, shifting the focus towards a particular flavour of models that include non-trivial interactions between the inflaton and the space-time metric \citep{Planck:2018jri}. Just as ambitious is the cosmic collider programme, which aims to use measurements of the non-Gaussianity of large scale structure to learn about the particle content in the inflationary era \citep{Arkani-Hamed:2015bza}. 

On a different front, there have been surprising results from attempting to measure the dark energy equation of state in more detail. Recent measurements of the expansion rate as a function of redshift, using distant supernovae or the imprint of baryon acoustic oscillations on the galaxy distribution, seem to indicate with mild statistical significance that accelerated expansion is driven by a dynamical field rather than a cosmological constant \citep{DESI:2024mwx}. This dynamic field is likewise envisioned as a scalar field and goes by the name \textit{quintessence} to distinguish it from inflation \citep{Caldwell:1997ii}. While the results are preliminary, these measurements have triggered a veritable gold rush in terms of model building. A vast number of possible explanations for this dynamical field, either new or existing \citep{Copeland:2006wr}, have been pored over in detail with the hope that we might be able to make definitive statements about the fundamental physics at play in the dark universe (e.g., \cite{DESI:2024kob, Wolf:2024eph, Wolf:2024stt, Ye:2024ywg, Wolf:2025jed}). The hope is that the recent detailed measurement of the (accelerated) expansion rate of the Universe can be used to determine the potential of the quintessence field or the specific form that new non-gravitational forces might take, or even bring to light new theories of gravity that might supersede general relativity. Such theories have the merit of leading to new signatures---modifications to gravitational collapse, new fifth forces, etc.---which might be probed by large scale structure and other means.

The case of dark matter is far richer and more complex, interweaving the methods of physical cosmology with direct astrophysical probes, from high energy astrophysics to galactic dynamics,  but also relying on direct detection and table top experiments. A huge range of fundamental physics models have been proposed as well as more prosaic, astrophysical models (involving compacts objects and black holes). While the effects of dark matter on the expansion of the Universe, and on the clustering of matter, are well described by a pressureless fluid, it is possible to distinguish between different broad classes of dark matter candidates through the effect they have on the substructure of galaxies and through their decay and annihilation signatures (which can be detected in, for example, $\gamma$-rays) arising in the regions of high density. In some sense, the case of dark matter aligns itself much more with more conventional methods for probing fundamental physics and does not rely exclusively on the tools of cosmology. 

The view that it should be possible to make meaningful statements about the microphysics of inflaton, dark matter, and dark energy persists to this day. Given the nature of cosmology's achievements, such a view is hardly surprising. The precision with which it is possible to measure cosmological parameters which are, to some extent, directly related to fundamental physics is remarkable. We can measure the energy density of the dark energy and dark matter to percent level accuracy, we are reaching a point where we can pin down the equation of state of the dark energy component to percent level (under some restrictive assumptions), and we already have measured the primordial anisotropies that (seem to have) originated from inflation to sub-percentage levels of precision. The rise of cosmology as a fundamental physics---as a ``particle accelerator in the sky''---has been inexorable and extremely successful. The success reflects itself in the number of papers written, the number of annual conferences on the topic, and how it has become core to the scientific cases for current and future facilities.

\section{The spectre of underdetermination}

There is no doubt that the story of physical cosmology is one of the resounding success stories of modern physics. A mathematical model of the Universe based on known physics has been shown to be consistent and predictive, in addition to being astonishingly simple and parsimonious. Yet it relies on the existence of three exotic components: inflation, dark matter, and dark energy. These components can be described completely in terms of their bulk, thermodynamic, or emergent properties. To be clear: just as we can describe water in terms of its density, pressure, viscosity, and other such properties, we can do the same for the unknown components in the case of cosmology. And from this bulk perspective, the model is complete.

However, the rise of cosmology as a fundamental science has set the bar higher. The goal is to determine the microphysics of the exotica: the underlying degrees of freedom, their equations of motion, and their actions which would supplement the standard model of particle physics. And that has driven an incredibly productive strand of theoretical physics, leading to the production of countless proposals. 
There are different opinions about what is the most mathematically consistent or theoretically correct approach to have; a purist might say that, ultimately, theoretical considerations will guide us to the unique correct solution, much as with the standard model of particle physics. A more pragmatic approach is that  the data will uniquely select the right answers.

But it is useful to revisit Withrow and Bondi's earlier debate. At that time, underlying the more general statements of Withrow and Bondi was the impasse between the Steady State and Big Bang cosmologies. Both were based on their own interpretations of the cosmological principle; and both were, to some extent, mathematically complete. It was a matter of opinion which version of the principle one chose to defend. 
Ultimately the situation was resolved unambiguously by the data. 

We now find ourselves with three open challenges: what are the microphysical underpinnings of inflation, dark matter, and dark energy?\footnote{To further muddy the waters, the case has been made that none of these components exist. For examples of the arguments deployed, see \cite{Turok:2002yq}, \cite{McGaugh:2014nsa} and \cite{Mohayaee:2021jzi}. If anything, this point of view further reinforces the issue being emphasized in this paper.} Will data resolve the situation as it did when cosmologists were debating between the Steady State and Big Bang models? Here, there are some serious reasons for doubt \citep{ferreira2021crossroads}. To see this, let us reiterate what we mean when we say ``cosmological data''. The cosmological data that has driven physical cosmology to its glorious heights are measurements of the expansion rate of the Universe over time (or redshifts) and measurements of the large scale structure of the Universe through various methods. These are the tools that are used by cosmology to place constraints on the models which one hopes will shed light on fundamental physics.

Let us focus first on the case of inflation.
Arguably inflation predicts that the Universe is flat, homogeneous, isotropic, and free of monopoles---but these predictions do not let us constrain inflationary models. For that, we look at the details of the primordial perturbations: the initial conditions created during inflation that lead to the large scale structure observed today. The specific characteristics of these initial conditions are generated during a narrow window of evolution of the inflaton field in which it traverses a short stretch of its potential. In other words, at most, we can probe a small part of the overall potential. 

We can infer the properties of this potential from a standard set of numbers that characterize the primordial perturbations: the overall amplitude, the spectral index, (possibly) one or two further numbers which describe how the spectral index changes with length scale, and the amplitude of primordial gravitational waves \citep{Planck:2018jri}. There is an even more ambitious view that it will be possible, using current and future observations, to constrain the spectral index of gravitational waves, and higher order moments or correlators that would let us charactertise the statistics of the primordial fluctuations beyond the usual assumption that they are Gaussian. Realistically, this would lead to another handful of numbers but which would be poorly constrained compared to the standard ones.

There are tremendous challenges to this programme. The amplitude of primordial gravitational waves is already constrained to be very small and, realistically, the expectation is to be able to improve these constraints by just over an order of magnitude \citep{CMB-S4:2016ple}. Non-cosmological signals such as the emission of gravitational waves due to our galaxy set an unavoidable threshold below which it is practically impossible to go. The primordial non-Gaussian signals will be obscured by the non-linear nature of gravitational collapse on smaller scales and will be limited by cosmic variance on the largest scales.\footnote{I.e.,\ the fact that one can only access a reduced number of statistically independent modes.} Furthermore, the role of systematic effects have been shown to play a large role on large scales: the galaxy affects how well one can measure the large scale distribution of mass and gas and correcting for it is challenging.

At the end of the day, one is left with a handful of well-measured numbers with which one hopes to pin down the model of inflation or, to be specific, the inflaton potential of which we probe only a small part. One can already see the threat of undertermination of theory by evidence. Over the decades, a large number of candidates for the theory of inflation have been proposed. An influential compilation of theories, the {\it Encyclopaedia Inflationaris} \citep{Martin:2013tda}, contains 118 such candidates (a recent update has 283 \citep{Martin:2024qnn}), many of which are viable. Furthermore it has been shown that it is straightforward to automatically generate large numbers of distinct microphysical theories which are compatible with existing data \citep{Sousa:2023unz, Wolf:2024lbf, Stein:2022cpk, Kallosh:2018zsi}. 

The situation with dark energy is similar. While inflation occurs at high energies, in the early Universe, and can be thought of a probe of ultraviolet physics, dark energy by contrast occurs at late time, low energies, and emerges in the infrared. To constrain the effects of dark energy with cosmological data, we rely on two types of observables. The first is direct (or indirect) probes of the expansion rate as a function of time or redshift. With constraints on the expansion rate, we can infer what is driving that expansion, and with good enough data we can disentangle the known contributions from the energy density which is driving the accelerated expansion. We can then begin to reconstruct the behaviour of this energy density and, for example, how its pressure is related to its energy density as a function of time---i.e.,\ its equation of state \citep{Planck:2015bue, DESI:2024mwx}.

Constraints on the equation of state have been improving gradually over time and it is hoped that with the next generation 
of surveys there will be significant further advances. If one assumes that the equation of state is constant, the goal is to measure it to within a percent. If one doesn't assume it is constant but allows for some parametrized form of evolution, constraints are obviously weaker but shed greater light on the nature of dark energy. Currently, the standard procedure is to consider two parameter fits for the equation of state but the hope is to be able to constrain more flexible fits with more parameters. Nevertheless, it is clear there is a trade-off between how well we can reconstruct the fine detail of the evolution of the equation of state (i.e.,\ how many parameters a fit has) and the accuracy of the reconstruction. 

The other tool for studying dark energy is the growth rate of structure: how fast gravitational collapse occurs as a function of time \citep{Ishak:2024jhs}. While this will be affected by the expansion rate of the Universe and therefore will be an additional probe of the equation of state, it will be more useful for probing other aspects of the dark energy---how it couples to gravity, whether the behaviour is different to that of a normal scalar field, etc. It will be particularly sensitive to whether the dark energy can be associated to, or mediate, new fifth forces in the Universe. And so it will enrich our understanding of dark energy. But, again, the amount of information  is limited to but a handful of numbers that will characterise the effect on gravitational collapse beyond what one might expect from our knowledge of general relativity.

We find ourselves in a situation which is entirely analogous to that which we described above for inflation. Over the past three decades, a veritable treasure trove of theories of dark energy has been constructed. The majority of the proposals involve scalar fields with potentials that come to dominate at late times. As with inflation, the cosmological observables will probe only a very short period in the evolution of the scalar field as it traverses a short stretch of its potential. Many different potentials will behave in a similar way over that short stretch, so in practice this means that even with the envisaged improvement in constraints on the equation of state or growth rate we will be left with a large family of models which are essentially observationally indistinguishable from the point of view of cosmological data \citep{Wolf:2023uno}.

Finally, the situation with dark matter is conceptually somewhat different, although, if one restricts oneself to the tools of physical cosmology, it is in practice not too dissimilar to the other two cases. As previously stated, the baseline is that dark matter is a non-relativistic particle which can be described effectively by a pressureless fluid. This minimal description is sufficient to explain essentially all cosmological observations: its impact on the expansion of the Universe and how gravitational collapse unfolds. Any particle which is sufficiently heavy and whose interaction cross section with other particles is sufficiently small will do the job. As one might expect, there are countless suitable proposals \citep{Bertone:2004pz}. Interestingly, the lack of experimental evidence for weakly interacting particles which might arise in supersymmetric theories has catalysed research into two distinct alternatives: ultra-light scalar fields and primordial black holes \citep{Bertone:2018krk}.  Depending upon the specific parameter choices for these models, they can lead to slightly different effects on the morphology of the cosmic web, although the effects are so slight that they are difficult to disentangle from other ingredients of the cosmological model (such as baryons and neutrinos) \citep{Chisari:2019tus}.

Taking stock of the situation regarding our three open problems, we have a set of a few numbers (e.g., constraints on the primordial power spectrum or the dark energy equation of state) which we can constrain to incredibly accurate precision and about which we expect to gain more useful empirical information in the future. This is a familiar epistemic reality in science that reflects \textit{transient} (or weak) underdetermination of the values/properties that these particular quantities possess; that is, underdetermination which could conceivably be broken in the future with more empirical information \citep{Duhem1954-DUHTAA, Sklar1975-SKLMC}. For example, we can plausibly envision a future in which we learn whether the dark energy equation of state is constant or dynamical, or similarly, a future in which improving constraints on the parameters that describe inflation eliminates certain classes of currently viable models while more clearly favoring others. However, the recent theoretical and observational developments sketched above seem to reveal a far more pernicious form of underdetermination in the background; a kind of \textit{permanent} underdetermination of the microphysical models that could reproduce the emergent numbers/properties to which we do have empirical access. Here, permanent underdetermination refers to the idea that there can be many distinct theories that, while technical empirically distinct from each other, produce empirical predictions which are arbitrarily close \citep{Pitts2010-PITPUF}. Considering that it now appears to be almost trivial to generate large numbers of microphysical models that are compatible with even the most optimistic constraints that can be placed on these numbers that describe the bulk/emergent properties of these exotic energy species, permanent underdetermination is an accurate descriptor for the spectre of underdetermination faced at the level of microphysical model building in modern cosmology.\footnote{For work discussing some different aspects of underdetermination in cosmology than those emphasized here, see \cite{Ellis2014-ELLOTP-4, Butterfield:2014twa, sep-cosmology}.} In a similar manner to how measuring the viscosity of a fluid will not uniquely pin down its microphysical structure, measuring the bulk properties of cosmological exotica will seemingly do no better.

\section{Statistical physics, spectroscopy and the birth of quantum mechanics}

In 1905---his \emph{annus mirabilis}---\cite{einstein1905brownian} wrote a paper, ``Investigations on the Theory of Brownian Motion'', in which he attempted to explain Brownian motion: the random motions of plant pollen particles which had been observed and documented by Robert Brown in the early 19\textsuperscript{th} century. For his explanation, which involved a clear mathematical description of the process, Einstein repurposed the blossoming field of the kinetic theory of gases which had been created by James Clerk Maxwell, Ludwig Boltzman and Josiah Willard Gibbs, and which proposed a microphysical explanation for the laws (and processes) of thermodynamics.

The kinetic theory of gases had emerged from the idea that the world is made up of fundamental constituents---atoms and molecules---and that their collective behaviour leads to bulk material properties which are observed. The atomist point of view had its critics (notable examples were Ernst Mach and Wilhelm Ostwald \citep{lindley2001boltzmann}), but once adopted offered a microphysical theory which could be used to derive the properties of, for example, gases and other macroscopic phenomena.

Einstein embraced the role of molecules and, using the kinetic theory of liquids, he derived a diffusion equation that would be obeyed by an ensemble of pollen particles immersed in water. He was able calculate a diffusion constant for the system and relate it to the mean distance travelled by the particles as a function of time. Conversely, a by-product of his analysis was that he was able to use measurements of pollen displacement to infer the size of water molecules. In other words, assuming that molecules existed, he could use observations of Brownian motion to constrain at least one of their properties.

Einstein's paper was concurrent with a growing tide of experimental results corroborating the atomistic view of the world. His work on Brownian motion was seen as indirect, but compelling, confirmation for the existence of atoms and molecules. And it came with a partial characterisation of these molecules, i.e., their size. But, on their own, the theory and observation of Brownian motion could not be used to learn more about the actual microphysics of how atoms and molecules were constituted. It was simply too blunt an instrument.

There were a number of different experimental results behind the consolidation of the atomic world view, from the discovery of the electron and the nucleus, the photo-electric effect, etc.---but the main driving force was the field of spectroscopy. From the early 19\textsuperscript{th} century with the discovery of spectral lines, a powerful tool for characterizing the structure of matter was developed such that, by the beginning of the 20\textsuperscript{th} century, there was an abundance of precise data and well-established experimental facts with which to interrogate the inner structure of the atom.

It took a while for a correct microphysical understanding of atoms and molecules to emerge. Rudimentary phenomenological models such as Rutherford's plum pudding model were first invoked, followed by attempts to modify the classical laws of physics with selection rules in Bohr's atom, which established what later became known as the ``old'' quantum mechanics. It was only when Heisenberg, rooting himself in the view that spectroscopy had given him of the microscopic world, developed matrix mechanics that the correct theory emerged. Heisenberg's formulation was shortly followed by Schr\"{o}dinger's wave mechanics, and the new quantum mechanics was thereafter established by Dirac and von Neumann.\footnote{For a philosophical discussion of this episode from the point of view of theory unification and equivalence, see \cite{Muller, Muller2}.}
While Einstein's theory of Brownian motion had shone a light on the microscopic world and supplied a tool for quantifying some of its properties, it would have been completely ineffectual at helping establish quantum mechanics and atomic structure: that was consolidated by spectroscopy.

Spectroscopy had a number of powerful features. To begin with, it is remarkably precise, with a number of accurately measured lines at particularly frequencies or wavelengths. Furthermore, for each system, there was an abundance of lines leading to, in effect, a physical barcode or fingerprint for a particular atom or molecule. This could be done for a wide range of different substances so much so that spectroscopy became an activity (and profession) in and of itself: a way to map out the microscopic world in unbridled detail. And finally, it was possible to use spectra to study different physical effects on substances, such as the role that electric and magnetic fields might play. All of this meant that spectroscopy was a multifaceted, flexible, and precise way in which to probe the fundamental constituents of matter.

One can think of physical cosmology relative to fundamental physics in the same way as one can think of Brownian motion relative to atomic physics and quantum mechanics. We have an incredibly powerful mathematical approach rooted in known physics which allows us to use measurements of the expansion rate and how galaxies are distributed and clustered to infer properties belonging to the Universe: the type of components of which it is made, how many of them there are, and how they evolve with time. This is analogous to Einstein figuring out, from the distance grains of pollen travel as function of time, the size of water molecules. What we can't do is say something more fundamental about these components---i.e., their microphysical properties. In the same way, we cannot use Brownian motion to reconstruct the energy levels of the atom, or measurements of viscosity to reconstruct the molecular composition of a fluid. Although with the current and future generations of cosmological data we will make measurements with ever-increasing precision, there is a fundamental limit to what we can leverage with this data.

Returning to Einstein's methodological distinction between phenomenological physics and ``atomistic thought'',
\citet[pp.\ 370--1]{Einstein1936} famously wrote the following of his own field equations:
\begin{quote}
The resulting
disappearance of the divergence of the right side produces the
``equations of motion'' of matter, in the form of partial differential equations for the case where $T_{ik}$ introduces, for the description of the matter, only \emph{four} further independent functions
(for instance, density, pressure, and velocity components [...])

By this formulation one reduces the whole mechanics of gravitation to the solution of a single system of covariant partial differential equations. [...] 
It is sufficient---as far as we know---for the representation of the observed facts of celestial mechanics. But it is similar to a building, one wing of which is made of fine marble (left part of the equation), but the other wing of which is built
of low-grade wood (right side of equation). The phenomenological representation of matter is, in fact, only a crude substitute for a representation which would do justice to all known
properties of matter.
\end{quote}
Here, Einstein explicitly points to shortcomings of the stress-energy tensor (encoding pressure, density, etc.)\ as a phenomenological construct, ``low-grade wood'' in his own words. The question is: under what circumstances might we be able to do better than such a phenomenological description, in the case of modern cosmology?

\section{Prospects for a cosmological spectroscopy}

One might then ask: what is the equivalent of spectroscopy for physical cosmology? What alternative might one use to probe deeper into the fundamental microphysical structure of inflation, dark matter and dark energy? The fundamental problem is, of course, the energy, mass, or length scales involved. Starting with inflation, one would need to have some process by which we could explore physics at ultra-high energies. Most inflationary proposals are, roughly, on scales of $10^{17}$ GeV although there are some at higher and lower scales. Thus, it is inconceivable that we will be able to reach them with current (or even future) accelerator technology. 

The observables which we use to pin down inflation probe the part of the potential which was traversed by the inflaton field around fifty to sixty e-foldings before the end of inflation.\footnote{I.e.,\ when the scale factor was $e^{60}$ time smaller than it was at the end of inflation} At the end of inflation, the Universe has to reheat and generate the particle species that are consistent with what is observed today. One could imagine that, looking at the resulting particle abundances, one might be able to glean more information about the inflaton potential in an altogether different regime than that probed by the CMB and large scale structure. Unfortunately, this is not the case. In order to reheat the Universe, the inflaton has to couple to other particles, greatly increasing the number of possibilities for how it might be integrated with the standard model. The observables that one might use to constrain particle production at the end of inflation are practically non-existent, and hence even less useful for pinning down the inflationary model.

Other possible probes are being explored. One is to use gravitational wave detectors to measure the stochastic gravitational wave background generated during the inflationary period. Another is to measure spectral distortions in the CMB which would be generated by very small scale fluctuations dissipated during the radiation era and around recombination. Both of these probes have the merit of being sensitive to either different types of fluctuations or fluctuations generated at a different time (and therefore a different part of the potential) to that probed by large scale structure. But the amount of information gleaned from such measurements is again remarkably small and not necessarily very precise. It is implausible that such measurements would be able to greatly reduce the range of models allowed by current and future large scale structure measurements. In particular, none of these approaches yield anywhere near the abundance of data that spectroscopy brought to atomic physics.

In the case of dark energy, the energy scales are very low: of order $10^{-3}$ eV corresponding to length scales of order the horizon, or tens of gigaparsecs. The only structure large enough to probe these scales is the entire Universe---and we are already doing that with physical cosmology. There is no known particle physics process, or other gravitational process, that allows us to access these energy or length scales. And similarly to inflation, if dark energy is driven by some dynamical field, the observables used to constrain its properties probe only a very narrow range of the field's evolution. Essentially, this is because constraints on the dark energy equation of state indicate that it is close to the cosmological constant value; thus, any evolution that the field might have undergone is constrained to be very small.
This leaves a huge number of distinct possibilities that will be indistinguishable over this very short range of field evolution.

There is the possibility that dark energy arises from some modification to general relativity on large scales which results in new fifth forces that might be probed by other means \citep{Ferreira:2019xrr}. But we already have remarkably tight constraints on small scales \citep{Adelberger:2003zx} which means that there must be some mechanism that inhibits the fifth forces on small scales \citep{Babichev:2013usa,Burrage:2017qrf}. One can then look for the smoking gun for this mechanism by looking for signatures in objects of intermediate size  such as galaxies and clusters \citep{Baker:2019gxo}. While there have been some attempts to identify such signatures, at most all one can do is rule out particular theories. Moreover, not all theories will have observable signatures that can be used to constrain them. So, unless the model of dark energy has some novel gravitational effect and that effect is detected, there is little prospect of winnowing the space of theories down to one which is preferred on observational grounds. 

There is somewhat more occasion for optimism in the case of dark matter as there has been a concerted campaign to look for non-cosmological evidence of dark matter particles, and these different approaches may be able to more directly access the microphysical nature of the dark matter candidates. Leading these efforts have been direct detection experiments which have slowly but systematically been sweeping through the parameter space of weakly interacting massive particles (WIMPs). While there have been some claims of a detection by one experiment \citep{DAMA:2008jlt}, they have not been corroborated by other experiments and the consensus is that there is as yet no experimental evidence for supersymmetric WIMPs \citep{LZ:2022lsv}. The technology is now sufficiently developed that current experiments will reach the limit of sensitivity set by the neutrino floor below which the neutrino background obscures any possible WIMP detection. Unfortunately, there is still a large swathe of parameter space for WIMP models which would be allowed.

An alternative approach is to look at methods of indirect detection of dark matter by studying high energy emission at various frequencies from stars, black holes or putative concentrations of dark matter at the center of galaxies or clusters of galaxies \citep{Bertone:2018krk}. Depending upon which structures are being observed, one can hone in on particular mass ranges or cross sections, much as with accelerator physics and---as with direct detection---the nature of the constraints are more similar to spectroscopy than those obtained on cosmological scales. Nevertheless, they do allow for significant amount of underdetermination: one has effectively, the ability to constrain a very limited number of parameters (or couplings) over a narrow range.

The interest in alternatives to WIMPs has also led to novel, non-cosmological probes. A notable example is the axion or, more generally, light scalar fields, originally postulated to resolve the strong CP problem. These particles  lead to different effects in different regimes. For example, they might accrete around black holes, thereby affecting the gravitational wave signals emitted during binary black hole mergers \citep{Aurrekoetxea:2024cqd}. Or they might accumulate around rotating black holes, leading to high energy emission \citep{Brito:2015oca}. A new wave of experiments are being developed, using recently developed techniques in quantum technology to look for specific signatures of how these scalar fields might interact with the rest of the world \citep{Badurina:2019hst}. Again, the models probed and the range of parameters are limited. 

The case of dark matter highlights an interesting aspect of this kind of research: the `streetlight effect'. There are limited windows into dark matter candidates, and only in particular regimes are they amenable to non-cosmological probes. So resources are, inevitably, focused on these probes. It is analogous to the driver who has lost their car keys on a dark street and decides to look only under the lamppost. There is no compelling reason for them to be there, but it is the only place where they can effectively look. In the case of dark matter, the same process is unfolding and if there is no detection with these limited number of probes, this might not shed much light on the far vaster space of unexplored models and regimes.\footnote{As far as we're aware---and perhaps surprisingly---there is as-yet no systemic study of the streetlight effect in the philosophy of science. And yet, there are good grounds to think that it is quite ubiquitous in modern science: for another example in the context of phosphoproteomics in modern biology, see \citep{Sharma2021-SHAHSI-3}.}

\section{Are we back where started?}
 
Physical cosmology is quite clearly a science using any sensible criterion which one might wish to invoke. 
Given cosmology's consistent, predictive success in uncovering novel facets of the nature of the Universe, what else could one reasonably demand from a science? Whitrow would be unable to deny physical cosmology its rightful place. However, where the answer becomes more nuanced, if we were to follow Whitrow's guiding principle, is if we try to assess the status of \emph{fundamental} cosmology, where one envisions uncovering the microphysical explanation for {\it all} the processes that underpin the cosmological model. Proposals abound for the microphysics but there is no clear path to finding the unique explanation.  As we've seen, it would seem that this particular aspect of cosmology is, potentially, permanently underdetermined.

Thus it behooves us to step back and speculate at how cosmology might evolve from here. In this article, we have focussed on observational methods of physical cosmology which, we reiterate, involve constraining the expansion rate of the Universe over time and the growth and morphology of large scale structure throughout time and space. It is with these methods (and these methods alone) that we have been able to establish $\Lambda$CDM or one of its close variants (where $\Lambda$ is replaced by dynamical dark energy, characterized in terms of its equation of state) as a remarkably predictive model.

The observational methods  of physical cosmology are constantly being improved and refined as we increase the quality and quantity of data. As a result, we expect to get ever more precise constraints on the parameters of the $\Lambda$CDM model and its close variants. There is no doubt that one should continue down this path, strengthening our knowledge of the model and testing its predictive power, but is it clear when we should stop \citep{2021NatAs...5..855L}?

As constraints improve, we might begin to see cracks or inconsistencies in the model. Indeed, we already have.  Two particular inconsistencies, or ``tensions'', have been the subject of much speculation and analysis \citep{Verde:2019ivm, DiValentino:2021izs}. The first is the ``Hubble tension''. If one takes the $\Lambda$CDM model and fixes its parameters to be consistent with measurements of anisotropies of the cosmic microwave background, it predicts an expansion rate today, or Hubble constant, which is consistently lower than if one attempts to constrain the Hubble constant directly by measuring the recession of galaxies. This tension is often characterized as an inconsistency between early time constraints (the CMB) and late time constraints (the recession of galaxies) on $\Lambda$CDM. 

The second tension, known as the ``$S_8$ tension'', can be formulated in a similar way. In this case, it is the amplitude of clustering in the Universe (i.e.,\ the amount of gravitational collapse that has occured) from early time constraints (again, from the CMB) which is different to the one inferred from late time constraints (now from the clustering galaxies inferred from weak lensing). The statistical significance is lower than the Hubble tension, but it is nevertheless a source of speculation. 

In both of these examples, the data is pointing towards an inconsistency in the mathematical model. It may be that these tensions are a transient phenomenon and that, with better data, they can be resolved or will otherwise go away. But if they persist, then $\Lambda$CDM may need to be modified. Inevitably, such modifications will have to be small, given how successful $\Lambda$CDM is at explaining so many other observations. But the modifications may give us deeper insights into what might underpin the phenomenological pillars of $\Lambda$CDM.

Another possibility is that other, non-empirical, criteria might single out particular microphysical models. In particular, solutions which are parsimonious, in the sense that they repurpose {\it existing} aspects of the cosmological model to explain any of the three exotic phenomena, might be favoured. A few examples come to mind. In the case of inflation, one possibility is that the Higgs field, a core element of the standard model of particle physics, might also be responsible for inflation \citep{Bezrukov:2010jz}. Such a theory makes very specific predictions for the amplitude of the stochastic gravitational wave background that is within range of sensitivity forecasts for the next generation of CMB experiments. In the case of late time acceleration, it has been argued that the non-linearities of general relativity might play a role in modifying the equations that model the expansion rate. The proponents of what is called the ``backreaction'' model argue that the simplifying assumption of homogeneity on the largest scales has been applied incorrectly and that the large scale structures that have been observed in maps of the distribution of galaxies will lead to modifications which, in turn, can be interpreted as a new source of energy driving the expansion of the Universe \citep{Buchert:2011sx}. However, this is a fringe view and very difficult to flesh out mathematically in detail; approximate estimates of the magnitude of the backreaction are as of now inconclusive.\footnote{While backreaction is no doubt the most ontologically parsimonious proposal for dark energy, the syntactic/mathematical complexity is far greater which muddies the water with a non-trivial trade-off between ontological and syntactic parsimony \citep{Wolf:2024aeu}.}
In the case of dark matter, examples include the QCD axion (which in turn is a proposal to solve the strong CP problem) \citep{Sheridan:2024vtt} or astrophysical black holes (which have already been detected). 

If any one of the above-described scenarios were to work out, one could reasonably leverage non-empirical arguments in favor of these proposals over their underdetermined alternatives. For example, successfully attaching inflation, dark matter, or dark energy solutions to well-confirmed and established aspects of general relativity or the standard model of particle physics might allow one to make a meta-inductive argument in their favor, whereby one uses the prior empirical successes of the established research programmes to infer additional meta/non-empirical support for such proposals \citep{Dawid2013-DAWSTA}. Another possibility is to argue that there is an epistemic dimension to the theory virtues at play here (such as parsimony or coherence), meaning that their presence gives us further warranted confidence in the merits of the proposals that possess them \citep{Schindler2018-SCHTVI-5}. Nevertheless, even under one of these optimistic scenarios, room for speculation will persist. Any number of other, less economical candidate explanations are still possible. Unfortunately, one would descend back into Whitrow's undesirable realm of opinions and ambiguity, particularly if one subscribes to the idea that such theory virtues are only reflective of pragmatics or matters of convenience \citep{VanFraassenBas1980-VANTSI}.

A more systematic approach would be to embrace the fact that each of the three exotica emerge at particular energy or length scales and only consider effective models that describe the physical processes at those energy scales. To some extent, the $\Lambda$CDM model is exactly that. But one can go further and deploy the ``effective field theory'' approach which has gained traction in fundamental physics since it was first proposed in the late 1970s \citep{Weinberg:1991um}. There, one assumes a certain fundamental field content (for example scalar fields) and certain symmetries and then proceeds to construct the most general action valid at a particular energy scales \citep{Burgess:2020tbq}. The idea then is that any microphysical theory with that field content and those symmetries will inevitably look like that effective field theory action at those energy scales. Such an approach has been proposed for inflation \citep{Weinberg_2008} and dark energy \citep{Park_2010},\footnote{This approach is not to be confused with the effective field theory approach that studies the symmetries of linear perturbations, with inflaton or quintessence field content, on an arbitrary, homogeneous, expanding background \citep{Cheung:2007st,Gubitosi:2012hu}.} where it has focussed primarily on single scalar field theories. The effective field theory of dark matter has been mostly applied to direct detection \citep{Fitzpatrick:2012ix} and at the Large Hadron Collider, focussing on WIMP-like particles,\footnote{Again, this approach is not to be confused with attempts at constructing an effective field theory approach to non-linear perturbation in large scale structure \citep{Carrasco:2012cv}.} but there is no approach which can cover the whole diversity of models that have been proposed for cosmology. 

Effective field theory is promising and may lead to the most unambiguous constraints on microphysics from cosmology. It is, nevertheless, still at the mercy of particular choices, i.e., the field content and the symmetries, and that can, again, lead to underdetermination. This degree of underdetermination in cosmology may then be no different from the underdetermination one has in the standard model of particle physics and how it can be used to constrain theories of grand unification.

We should of course be mindful of history and how situations which might have seemed inevitably underdetermined did ultimately progress rapidly. In the early 19$^{\rm th}$ century, August Comte---the mastermind behind the positivist school of philosophy---wrote in his {\it Cours de Philosophie Positive} \citep{comte1830cours}:
\begin{quote}
On the subject of stars, all investigations which are not ultimately reducible to simple visual observations are [...] necessarily denied to us. While we can conceive of the possibility of determining their shapes, their sizes, and their motions, we shall never be able by any means to study their chemical composition or their mineralogical structure [...] 
I regard any notion concerning the true mean temperature of the various stars as forever denied to us.
\end{quote}
It wasn't long before Gustav Kirchoff had shown how spectra could be used to determine the chemical composition of an object and stellar astrophysics was born.

As was the case with Kirchoff, one might speculate that the equivalent to spectroscopy might play similar role in cosmology. A new form of experimental or astronomical observation might arise which would supply a completely different and varied perspective on the microphysical processes that affect cosmological evolution. While we can be optimistic in the case of dark matter, where energy scales are potentially accessible through experimental or astrophysical probes, it seems inconceivable to reach the very high or very low energy scales required to pin down inflation or dark energy (or any alternatives to them which might be considered). 

Eugene Wigner, in a series of recollections \citep{szanton1992recollections}, described the atmosphere in theoretical physics during his stay in Albert Einstein's group in Berlin: 
\begin{quote}
Until 1925, most great physicists, including Einstein and Max Planck, had doubted that man could truly grasp the deepest implications of quantum theory. They really felt that man might be too stupid to properly describe quantum phenomena. [...] 
``Is the human mind gifted enough to extend physics into the microscopic domain---to atoms, molecules, nuclei, and electrons?''
Many of those great men doubted that it could.
 \end{quote}
As we know, spectroscopy, Werner Heisenberg, and Erwin Schr\"{o}dinger went on to transform physics. It's possible---and, indeed, we very much hope!---that the view put forward in this article regarding the fundamental limitations of microphysical modelling in modern cosmology is just as shortsighted as of those in Einstein's circle before 1925, and that a new conceptual revolution in cosmology will resolve these open problems in one fell swoop. While it's very difficult for us to envisage how that will be the case (given that inflation, dark matter, and dark energy all play out on such different energy scales), these misgiving of course should not deter us from trying.

\medskip

\textit{Acknowledgments.---}
PGF thanks the Royal Astronomical Society for being given the opportunity to present these ideas in the 2022 Gerald Whitrow Lecture. We are very grateful to D. Alonso, H. R. Brown, C. Burgess, G. Efstathiou, E. Linder, J. Peacock, J. Peebles, J. Silk and A. Wishart for helpful discussions. PGF is supported by STFC and the Beecroft Trust. WJW is supported by the HAPP center at St. Cross College, University of Oxford and the British Society for the Philosophy of Science.

\bibliographystyle{elsarticle-harv} 
\bibliography{refs}

\end{document}